\documentclass[letter]{aa} 

\usepackage{graphicx}

\usepackage{lscape}
\usepackage{amssymb}
\usepackage{amsmath}

\usepackage{txfonts}

\usepackage{booktabs}
\usepackage{subfigure}
\usepackage{placeins}
\usepackage{verbatim}
\usepackage{adjustbox}
\usepackage[toc,page]{appendix} 
\usepackage{tabularx}
\usepackage{lscape}
\usepackage{textcomp}
\usepackage{gensymb}
\usepackage{multicol}
\usepackage{multirow}
\usepackage[symbol]{footmisc}
\usepackage{tikz}
\usepackage{caption}
\usepackage{longtable}

\usepackage{natbib}
\usepackage{supertabular}

\usepackage{color}
\usepackage{lscape}

\usepackage{tablefootnote}
\include{shortcuts}

\usepackage{hyperref}  
\hypersetup{colorlinks=true,linkcolor=[rgb]{1.,0.2,0.2},citecolor=[rgb]{0.1,0.4,1.},filecolor=[rgb]{0.7,0.2,0.2},urlcolor=[rgb]{0.7,0.2,0.2}}

\def\macs1149{MACS 1149}

\def\NHUNIT{\ifmmode {\rm \,cm^{-2}} \else $\rm \,cm^{-2}$ \fi} 

\def\aap{{A\&A}}

\def\apjs{{ApJ Supp.}}
\def\muKcmb{\ifmmode \,\mu$K$_{\rm CMB}$\else \,$\mu$K$_{\rm CMB}$\fi}
\newcommand{\planck}{\Planck}

\newcommand{\OmegaM}{\ifmmode\Omega_{\rm M}\else $\Omega_{\rm M}$\fi}

\newcommand{\thetaMC}{\theta_{\rm MC}}

\providecommand{\Planck}{\textit{Planck}}
\providecommand{\planck}{\Planck}

\providecommand{\text}[1]{\rm{#1}}

\newcommand{\begm}{\begin{pmatrix}}
\newcommand{\enm}{\end{pmatrix}}

\def\pmb#1{\setbox0=\hbox{#1}%
    \kern-.025em\copy0\kern-\wd0
    \kern.05em\copy0\kern-\wd0
    \kern-.025em\raise.0433em\box0}

\def\p2Y{\;_2Y}
\def\m2Y{\;_{-2}Y}

\newcommand{\mksym}[1]{\ifmmode {\rm #1}\else #1\fi}

\providecommand{\text}[1]{\rm{#1}}

\newcommand{\bd}{\begin{displaymath}}
\newcommand{\ed}{\end{displaymath}}
\newcommand{\be}{\begin{equation}}
\newcommand{\ee}{\end{equation}}
\newcommand{\beaa}{\begin{eqnarray*}}
\newcommand{\eeaa}{\end{eqnarray*}}
\newcommand{\bea}{\begin{eqnarray}}
\newcommand{\eea}{\end{eqnarray}}

\newcommand{\cosmomc}{{\tt cosmomc}}
\newcommand{\camb}{{\tt camb}}

\begin{document} 

   \title{Cosmography with Supernova Refsdal through time-delay cluster lensing: 
independent measurements of the Hubble constant and geometry of the Universe}

\author{
C.~Grillo \inst{\ref{unimi},\ref{inafmilano}} \and
L.~Pagano \inst{\ref{unife},\ref{infnfe},\ref{ias}}\and
P.~Rosati \inst{\ref{unife},\ref{inafbo}} \and
S.~H.~Suyu \inst{\ref{tum}, \ref{mpa}, \ref{asiaa}}
}
\institute{
Dipartimento di Fisica, Universit\`a  degli Studi di Milano, via Celoria 16, I-20133 Milano, Italy \label{unimi}
\and
INAF -- IASF Milano, via A. Corti 12, I-20133 Milano, Italy \label{inafmilano}
\and
Dipartimento di Fisica e Scienze della Terra, Universit\`a degli Studi di Ferrara, via Saragat 1, I-44122 Ferrara, Italy \label{unife}
\and 
INFN -- Sezione di Ferrara, Via Saragat 1, I-44122 Ferrara, Italy \label{infnfe}
\and
Université Paris-Saclay, CNRS, Institut d’Astrophysique Spatiale, F-91405 Orsay, France \label{ias}
\and
INAF -- OAS, Osservatorio di Astrofisica e Scienza dello Spazio di Bologna, via Gobetti 93/3, I-40129 Bologna, Italy \label{inafbo} 
\and
Technical University of Munich, TUM School of Natural Sciences, Department of Physics, James-Franck-Str.~1, 85748 Garching, Germany \label{tum}
\and
Max-Planck-Institut f\"ur Astrophysik, Karl-Schwarzschild-Str.~1, D-85748 Garching, Germany \label{mpa}
\and
Academia Sinica Institute of Astronomy and Astrophysics (ASIAA), 11F of ASMAB, No.1, Section 4, Roosevelt Road, Taipei 10617, Taiwan \label{asiaa}
           }             


 
\abstract{We present new measurements of the values of the Hubble constant, matter density, dark energy density, and dark energy density equation-of-state parameters from a full strong lensing analysis of the observed positions of 89 multiple images and 4 measured time delays of SN Refsdal multiple images in the Hubble Frontier Fields galaxy cluster MACS J1149.5+2223. 
  By strictly following the identical modelling methodology as in our previous work, that was done before the time delays were available, our cosmographic measurements here are essentially blind based on the frozen procedure.
  Without using any priors from other cosmological experiments, in an open $w$CDM cosmological model, through our reference cluster mass model, we measure the following values: $H_0 = 65.1^{+3.5}_{-3.4}$~km~s$^{-1}$~Mpc$^{-1}$, $\Omega_{\rm DE}=0.76^{+0.15}_{-0.10}$, and $w=-0.92^{+0.15}_{-0.21}$ (at the 68.3\% confidence level). No other single cosmological probe is able to measure simultaneously all these parameters. Remarkably, our estimated values of the cosmological parameters, particularly $H_0$, are very robust and do not depend significantly on the assumed cosmological model and the cluster mass modelling details. The latter introduce systematic uncertainties on the values of $H_0$ and $w$ which are found largely subdominant compared to the statistical errors. The results of this study show that time delays in lens galaxy clusters, combined with extensive photometric and spectroscopic information, offers a novel and competitive cosmological tool.}

   
   \titlerunning{Cosmography with Supernova Refsdal through time-delay cluster lensing}
   \authorrunning{Grillo et al.}
   \maketitle

%

\section{Introduction}
\label{sec:intro}

In the standard cosmological model, the Universe is homogeneous and isotropic on its largest scales; its total mass-energy density is mainly in the form of dark-energy ($\Omega_{\Lambda} \approx 0.7$) and matter ($\Omega_{\rm m} \approx 0.3$, consisting of both ordinary and dark matter), and its geometry appears to be flat ($\Omega = \Omega_{\rm m} + \Omega_{\Lambda} \approx 1$). These values of the cosmological parameters imply a fairly recent transition from a decelerating to an accelerating cosmic expansion. The current expansion rate, the Hubble constant $H_0~\approx~70$~km~s$^{-1}$~Mpc$^{-1}$, represents another fundamental quantity that defines many of the most important scales in the Universe (e.g., its age and critical density). Despite the still unknown physical origin of the two dominant dark components, the standard cosmological $\Lambda$CDM model needs only a few parameters to reproduce well nearly all the current observational data: precision measurements of the temperature fluctuations in the cosmic microwave background (CMB; \citealt{Aghanim:2018eyx}), the observed abundances of light elements \citep{Cooke:2017cwo,Valerdi:2019beb,Izotov:2014fga,Aver:2015iza}, the large-scale distribution of galaxies (LSS; \citealt{Alam:2016hwk}), and the luminosity-distance relationship for distant type Ia supernovae (SNIa; \citealt{Scolnic:2017caz}). The estimates of different cosmological parameters from a single observational method are often correlated, exhibit significant uncertainties and can vary markedly with the adopted underlying cosmological model. This suggests that accurate and precise measurements of the cosmological parameters can only be obtained by combining complementary and independent techniques and, in fact, considerable efforts are still being made in this direction (see e.g. \citealt{mor22}).

In this paper, we focus on the Hubble Frontier Fields (HFF) galaxy cluster MACS J1149.5+2223 (hereafter MACS 1149; $z = 0.542$), where the strongly lensed SN “Refsdal” ($z = 1.489$) was discovered \citep{kel15,kel16a}. We exploit a large set of spectroscopically confirmed multiple images from different sources, identified from \emph{Hubble Space Telescope} deep imaging and Multi Unit Spectroscopic Explorer (MUSE) data (\citealt{tre16}; \citealt{gri16}, hereafter G16), and the measured values of the time delays of the SN Refsdal multiple images \citep{kel16,kel23a,rod16}. We build on our previous strong lensing analyses (G16; \citealt{gri18}, hereafter G18; \citealt{gri20}, hereafter G20), where we have described and tested extensively the methodology. We aim here at making the final, robust measurements of the values of the cosmological parameters introduced above, which are based on the data of MACS 1149 and SN Refsdal alone. A simultaneous measurement of all parameters in a very general cosmological model represents the most innovative part of our new time-delay cluster lensing probe, especially when compared to other methods/experiments.

\section{Methods}
\label{sec:methods}

To perform our strong-lensing and cosmographic study, we exploit the observed positions of 89 multiple images from 28 distinct background sources, distributed in redshift between 1.240 and 3.703 and approximated as point-like objects, and the measured values of the time delays of the images S2, S3, S4 and SX, relative to S1, of SN Refsdal, recently published by \citet{kel23a}. In particular, we adopt the median values and assume symmetric 1$\sigma$ errors for the time delays (obtained, respectively, from the 50th, and the semi-difference between the 84th and 16th percentiles of the `Combined' method in Table 15 of \citealt{kel23a}), i.e. ${\rm \Delta t_{\rm S2:S1}} = (9.9 \pm 4.0$) days, ${\rm \Delta t_{\rm S3:S1}} = (9.0 \pm 4.2$) days, ${\rm \Delta t_{\rm S4:S1}} = (20.3 \pm 6.4)$ days, and ${\rm \Delta t_{\rm SX:S1}} = (376.0 \pm 5.6)$ days. We model the cluster total mass distribution on the different cluster and member galaxy spatial scales with the combination of, respectively, 3 cored elliptical pseudo-isothermal and 300 dual pseudo-isothermal mass density profiles. All the details of this model have been presented in G16, G18 and G20 (r model). 

We emphasise this is the same mass parametrisation originally adopted in G16, which blindly predicted a position for SX less than $0.2$ arcsec away from the value observed later, with a time delay ${\rm \Delta t_{\rm SX:S1}}=361^{+20}_{-27}$ days, for an assumed value of $H_0 = 70$~km~s$^{-1}$~Mpc$^{-1}$ and a flat $\Lambda$CDM cosmological model with $\Omega_{\rm m}=0.3$. As in G18 and G20, we reoptimise here the same lens model by adding the constraints from the time delays and the position of SX, leaving all cosmological parameters free to vary. We remind that the values of the cosmological parameters affect the combinations of the angular-diameter distances appearing in the so-called time-delay distance, $D_{\Delta t}$, and family ratios, $\Xi$ (see Eqs. (2) and (3) in G18).
  
Following G20, we quantify possible systematic uncertainties on the measurement of the values of the cosmological parameters by also taking into account (1) a constant sheet of mass at the cluster redshift with a uniform prior between $-$0.2 and 0.2 on the value of the convergence k$_{0}$ (model labeled as $\kappa$) and (2) a more flexible power-law mass density profile (i.e., $\rho(r) \sim r^{-\gamma}$, in the outer regions $r \gg r_{\rm c}$) with a uniform prior between 1.4 and 2.6 on the value of $\gamma$ for the cluster main halo (model labeled as $\gamma$).

We consider the following four cosmological models:
\begin{itemize}
 \item[-] f$\Lambda$CDM: a flat ($\Omega_{\rm m}+\Omega_{\Lambda} = 1$) $\Lambda$CDM model;
 \item[-] o$\Lambda$CDM: an open $\Lambda$CDM model, in which the values of $\Omega_{\rm m}$ and $\Omega_{\Lambda}$ are independent;
 \item[-] f$w$CDM: a flat ($\Omega_{\rm m}+\Omega_{\rm DE} = 1$) $w$CDM model, in which the dark energy density is time dependent, with an equation-of-state parameter $w$;
 \item[-] o$w$CDM: an open $w$CDM model, in which the values of $\Omega_{\rm m}$ and $\Omega_{\rm DE}$ are independent and the dark energy density is time dependent, with an equation-of-state parameter $w$.
\end{itemize}
We adopt uniform priors on the values of $H_{0}$, between 20 and 120 km~s$^{-1}$ Mpc$^{-1}$, $\Omega_{\rm m}$, between 0 and 1, $\Omega_{\Lambda}$ or $\Omega_{\rm DE}$, between 0 and 1, and $w$, between $-2$ and $0$. We use the software {\sc Glee} \citep{SuyuHalkola10,SuyuEtal12} to perform the entire strong-lensing analysis.

\section{Results}
\label{res}

\begin{figure*}[ht!]
  \centering
  {\includegraphics[width=0.7\textwidth]{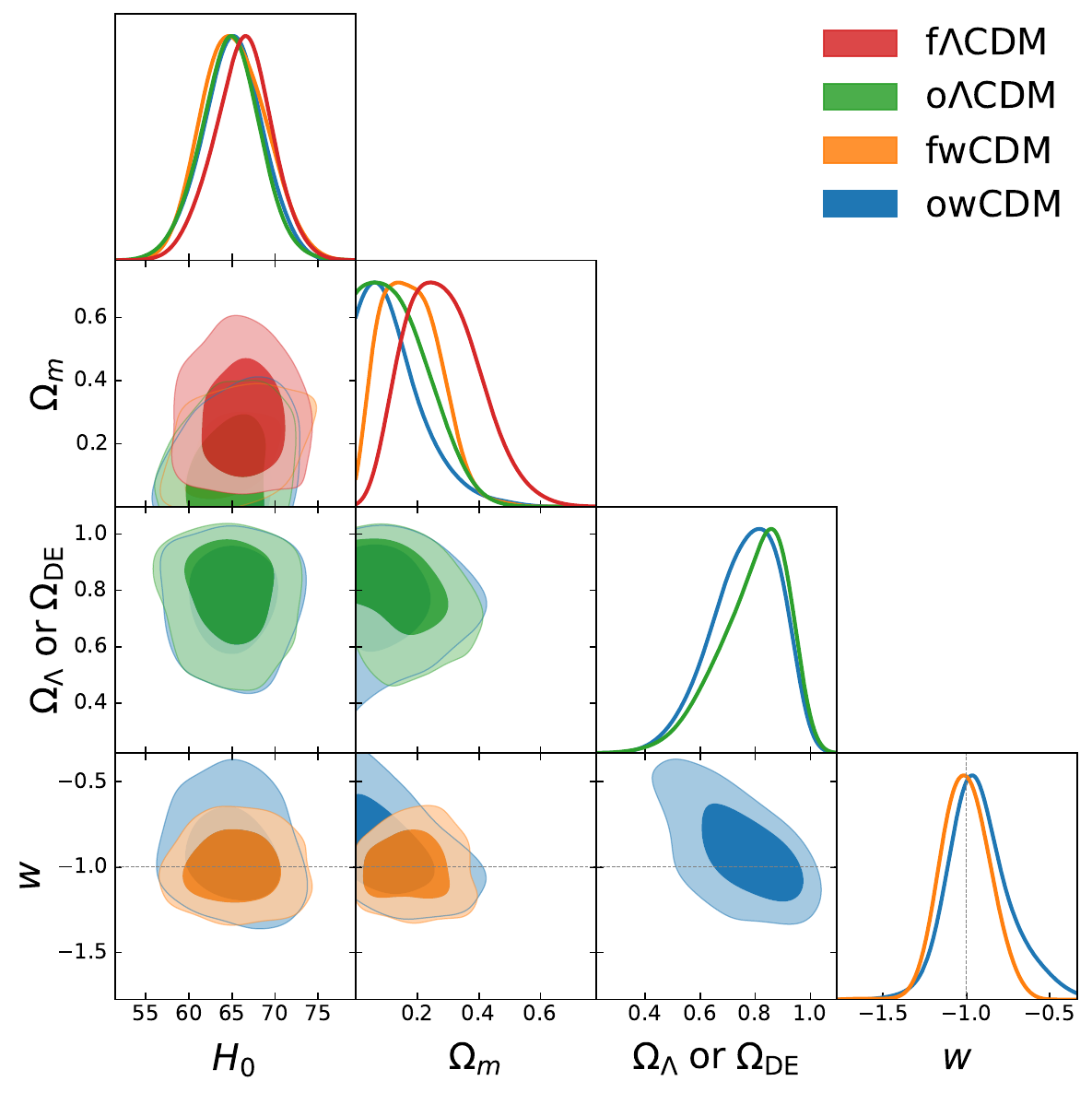}}
      \caption{Contour levels and marginalised probability distribution functions of $H_{0}$, $\Omega_{\rm m}$, $\Omega_{\Lambda}$ or $\Omega_{\rm DE}$, and $w$ (from top to bottom) for different cosmological models (in different colours) from time-delay cosmography in MACS 1149 only (r model). The contour levels on the planes represent the 68.3\% and 95.4\% confidence regions. Uniform priors on the values of all cosmological parameters ($H_{0} \in [20,120]$ km~s$^{-1}$~Mpc$^{-1}$, $\Omega_{\rm m} \in [0,1]$, $\Omega_{\Lambda}$ or $\Omega_{\rm DE} \in [0,1]$, $w \in [-2,0]$) are considered in our analysis. Final MCMC chains have $2 \times 10^{6}$ samples for each model. The dashed lines indicate models with $w=-1$, i.e. corresponding to a cosmological constant.}
         \label{fig:tri_all_models}
\end{figure*}

In all cosmological scenarios, our best-fitting models can reproduce the observed positions of the multiple images (SN Refsdal knots) with a rms offset of only 0.25$\arcsec$ (0.11$\arcsec$) and the measured time-delays within less than their 1$\sigma$ errors.

We summarise in Table \ref{tab1} the mean values and the 68.3\% confidence level (CL) intervals for $H_{0}$, $\Omega_{\rm m}$, $\Omega_{\Lambda}$ or $\Omega_{\rm DE}$, and $w$, measured in the different cosmological models from our strong lensing analysis of MACS 1149 and without any information from other cosmological probes. Chains are analysed using the \texttt{getdist} software \citep{Lewis:2019xzd}. Mean values are computed integrating the marginalised one-dimensional distributions. The confidence intervals are obtained slicing the marginalised distributions at constant ordinate and keeping the parameter intervals that contain 68.3\% of their probability. In Figure \ref{fig:tri_all_models}, we show the contour levels and the marginalised probability distribution functions of the same cosmological parameters and the corresponding 68.3\% and 95.4\% confidence regions.

\begin{table*}[h]
\caption{Cosmological parameters for different cosmological models from time-delay cosmography in MACS~1149 only (r model).}         
\label{tab1}      
\centering          
\begin{tabular}{c c c c c}    
  \hline\hline
  \noalign{\smallskip}
  Model & $H_0$ (km s$^{-1}$ Mpc$^{-1}$) & $\Omega_{\rm m}$ & $\Omega_{\Lambda}$ or $\Omega_{\rm DE}$ & $w$ \\
  \noalign{\smallskip}
\hline
\noalign{\smallskip}
f$\Lambda$CDM & $66.2^{+3.5}_{-3.2}$ & $0.28^{+0.10}_{-0.14}$ & $0.72^{+0.14}_{-0.10}$ & $\equiv -1$ \\
\noalign{\smallskip}
o$\Lambda$CDM & $64.8^{+3.5}_{-3.3}$ & $<0.34$ & $0.79^{+0.16}_{-0.09}$ & $\equiv -1$ \\
\noalign{\smallskip}
f$w$CDM & $65.3^{+3.5}_{-4.1}$ & $0.18^{+0.08}_{-0.11}$ & $0.82^{+0.12}_{-0.08}$ & $-1.00^{+0.14}_{-0.15}$ \\
\noalign{\smallskip}
o$w$CDM & $65.1^{+3.5}_{-3.4}$ & $<0.34$ & $0.76^{+0.15}_{-0.10}$ & $-0.92^{+0.15}_{-0.21}$ \\
\noalign{\smallskip}
\hline                  
\end{tabular}
  \begin{list}{}{}
\item[Notes --] Reported values are mean values, with errors corresponding to the 68.3\% confidence level (CL) intervals. In o$\Lambda$CDM and o$w$CDM models, for $\Omega_\textrm{m}$, 95.4\% upper limits are reported.
\end{list}
\end{table*}

First, we remark that the centres and the dispersions of the marginalised probability distribution functions (i.e., the mean values and the 68.3\% CL intervals) of $H_{0}$, $\Omega_{\Lambda}$ or $\Omega_{\rm DE}$, and $w$, depend only very mildly on the adopted cosmological model (i.e., flat or open, with a cosmological constant or a more general dark-energy component). The inference on the value of $\Omega_{\rm m}$ is instead quite sensitive to the chosen cosmology. 

In keeping with the results by G18 and G20, we find that we can estimate the value of $H_{0}$ with the smallest uncertainty, i.e. less than 6\%. The value of $\Omega_{\Lambda}$ or $\Omega_{\rm DE}$ is affected by an error ranging between 10\% and 20\%. Surprisingly, the value of $w$ can also be measured with an uncertainty slightly smaller than 20\%. The value of $\Omega_{\rm m}$ can be estimated with 40-50\% uncertainty in models with a cosmological constant and only upper limits can be obtained in more general dark-energy models (i.e., when $w$ is allowed to vary).

The achieved precision on the values of the considered cosmological parameters can be ascribed to the combined information contained in (1) the measured time delays of the multiple images of SN Refsdal, depending primarily on the value of $H_{0}$ and more mildly on those of the other parameters, and (2) the observed positions of the multiple images of the other sources at different redshifts, insensitive to the value of $H_{0}$ but sensitive to those of the other parameters (see G18 and G20 for more details).

\begin{figure*}[]
  \centering
  {\includegraphics[width=0.7\textwidth]{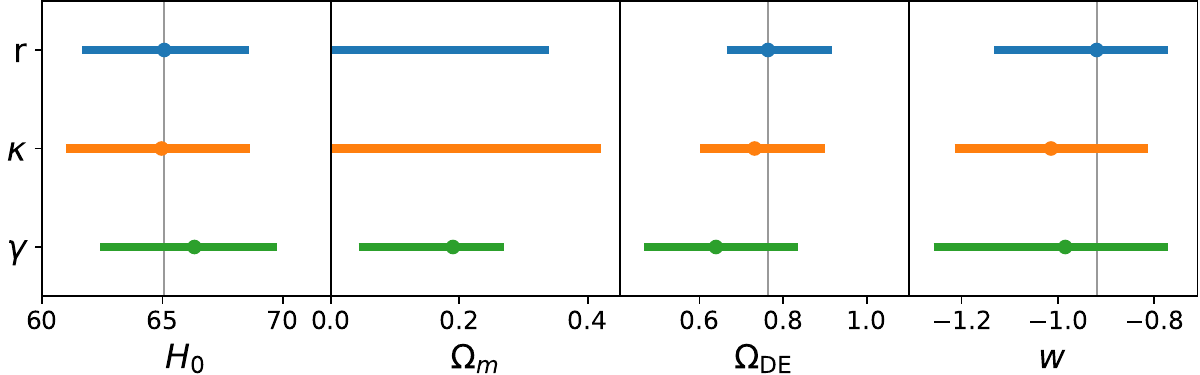}}
      \caption{Mean values and 68.3\% CL intervals of $H_{0}$, $\Omega_{\rm m}$, $\Omega_{\rm DE}$, and $w$ in an open $w$CDM (o$w$CDM) model for the reference model (r; see also Table~\ref{tab1}) and the extended models with an additional constant sheet of mass at the cluster redshift ($\kappa$) and a power-law density profile for the cluster main halo ($\gamma$). For $\Omega_{\rm m}$, in the r and $\kappa$ models, we show 95.4\% upper limits. Vertical lines highlight the mean values of the reference model.} 
         \label{fig:whisker_owcdm}
\end{figure*}

In Fig. \ref{fig:whisker_owcdm}, we show in the most general cosmological model considered in this study, i.e. an open $w$CDM model, the estimated values of the cosmological parameters for the three r, $\kappa$ and $\gamma$ cluster mass models. We remark that all measurements are consistent, with very similar uncertainties for the same cosmological parameter. The inferred value of $H_{0}$ is notably stable: the maximum difference between the three mean values (1.4 km s$^{-1}$ Mpc$^{-1}$) is approximately 40\% of the mean statistical error (3.6 km s$^{-1}$ Mpc$^{-1}$). The measured mean values of $\Omega_{\rm m}$ and $\Omega_{\rm DE}$ vary less than their average statistical uncertainties. The different mean values of $w$ span a range approximately equal to half the mean statistical error. By taking the differences in the mean values of the three models into account, we can include statistical plus systematic errors when quoting relative uncertainties for $H_{0}$ and $w$, and find approximately 6\% and 21\%, respectively. We have estimated the values of the Bayesian Information Criterion \citep[BIC;][]{schwarz78_BIC} and of the Akaike Information Criterion \citep[AIC;][]{akaike74_AIC} and have found positive evidence that the reference (r) model is preferred to the $\kappa$ and $\gamma$ models. We also notice the 68.3\% CL intervals for $\kappa$ and $\gamma$ include the fixed values adopted in the r model (i.e., $\kappa = 0$ and $\gamma = 2$).

\begin{figure*}[ht!]
  \centering
  {\includegraphics[width=0.8\textwidth]{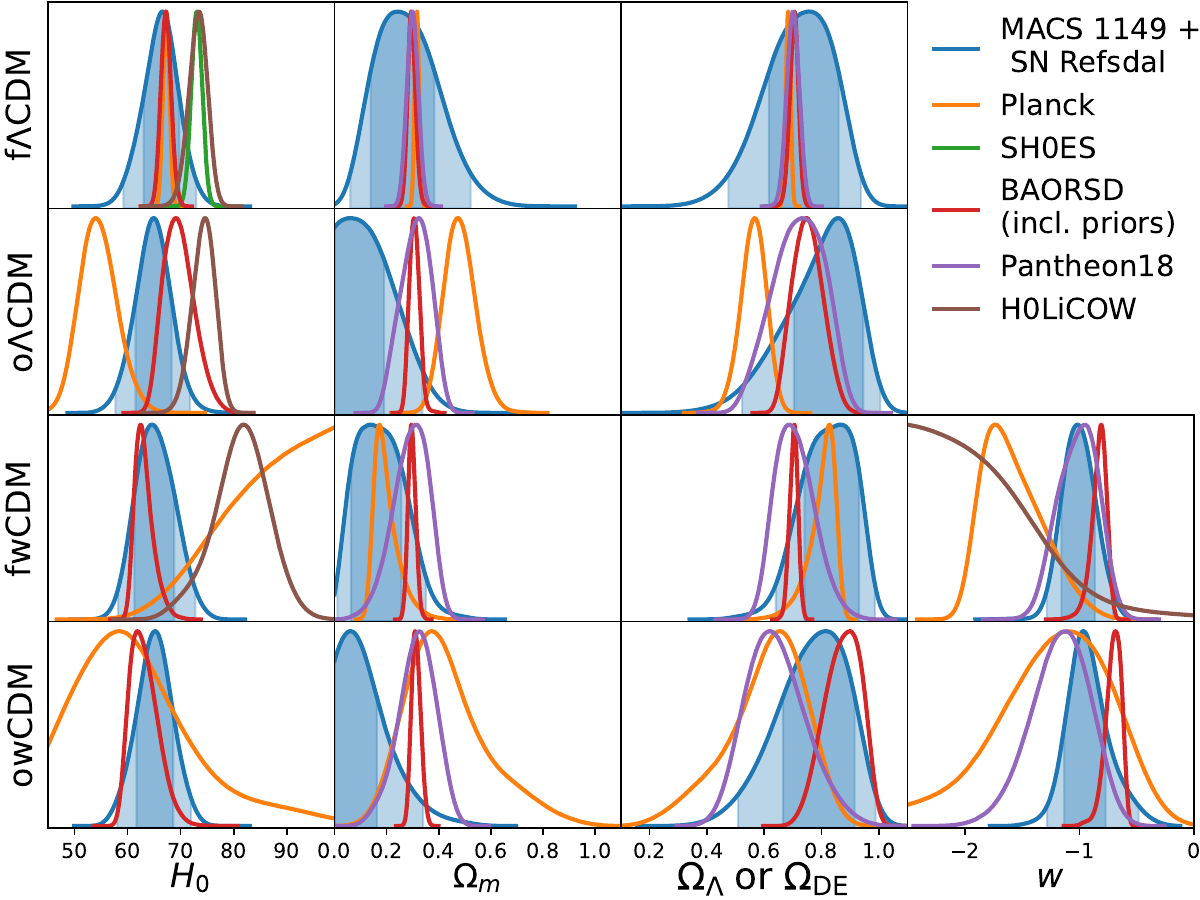}}
      \caption{Marginalised probability distribution functions of $H_{0}$, $\Omega_{\rm m}$, $\Omega_{\Lambda}$ or $\Omega_{\rm DE}$, and $w$ (from left to right) for different cosmological models (from top to bottom). Coloured lines show the one-dimensional posterior distributions obtained from various cosmological datasets: time-delay cosmography in  MACS~1149 (r model) in blue, Planck \citep{Aghanim:2018eyx} in orange, SH0ES \citep{rie21} in green, Baryon Acoustic Oscillations (BAO) and Redshift-Space Distortions (RSD) \citep{eBOSS:2020yzd} in red, the SN Pantheon sample \citep{Scolnic:2017caz} in purple, and H0LiCOW \citep{won19} in brown. Blue shaded regions represent the $68.3\%$ (darker) and $95.4\%$ (lighter) CL regions for time-delay cosmography in  MACS~1149. For SH0ES, only the probability density distribution function of $H_{0}$ in a f$\Lambda$CDM model is available and shown here. For BAORSD, we also include Gaussian priors on the values of $\omega_{\rm b}$ and $n_{\rm s}$ (see the list of datasets in Sect.~\ref{res}). For H0LiCOW, chains for the f$\Lambda$CDM, o$\Lambda$CDM and f$w$CDM models are available and here only the probability density distribution functions of $H_0$ and $w$ are shown, since the adopted smaller prior on $\Omega_{\rm m}$ causes a sharp edge in the posteriors of $\Omega_{\rm m}$ and $\Omega_{\Lambda}$ or $\Omega_{\rm DE}$.}
         \label{fig:plot_1d_all_models}
   \end{figure*}

\begin{figure}[!h]
  \centering
  {\includegraphics[width=0.45\textwidth]{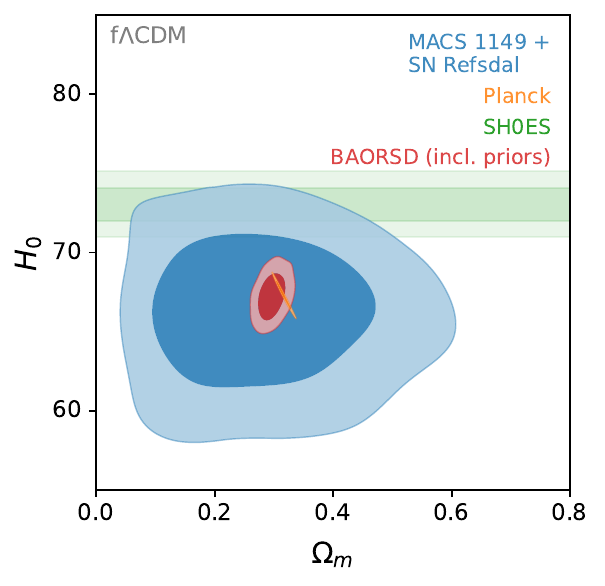}}
\caption{Contour levels, at the 68.3\% (darker) and 95.4\% (lighter) CL, of $H_0$ and $\Omega_{\rm m}$ in a flat $\Lambda$CDM (f$\Lambda$CDM) model for different cosmological probes: time-delay cosmography in MACS 1149 (r model; blue), Planck (orange), BAORSD (red). For completeness, we show also the SH0ES measurement of the Hubble parameter (green; \citealt{rie21}). BAORSD includes Gaussian priors on the values of $\omega_{\rm b}$ and $n_{\rm s}$. Details on these datasets are given in Sect.~\ref{res}.}
         \label{fig:plot_2d_Om_H0_lcdm}
\end{figure}

\begin{figure}[!h]
  \centering
  {\includegraphics[width=0.45\textwidth]{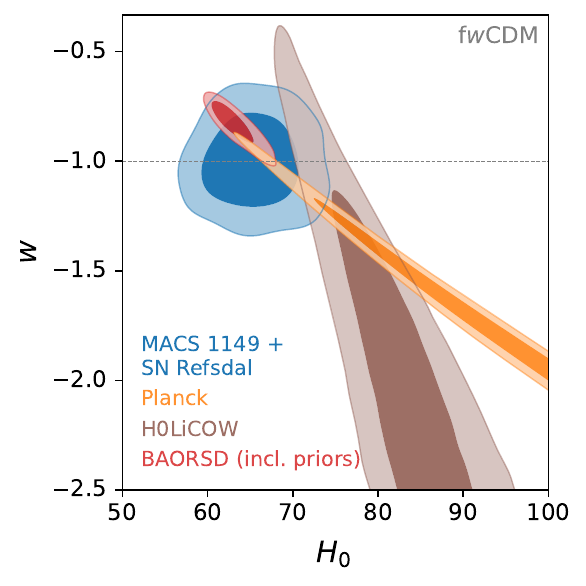}}
      \caption{Contour levels, at the 68.3\% (darker) and 95.4\% (lighter) CL, of $w$ and $H_0$ in a flat $w$CDM (f$w$CDM) model for different cosmological probes: time-delay cosmography in MACS 1149 (r model; blue), Planck (orange), H0LiCOW (brown) and BAORSD (red). The horizontal dashed line corresponds to models with $w=-1$, i.e. with a cosmological constant. BAORSD includes Gaussian priors on the values of $\omega_{\rm b}$ and $n_{\rm s}$. Details on these datasets are given in Sect. \ref{res}.}

         \label{fig:plot_2d_H0_w_wcdm}
\end{figure}

\begin{figure}[!h]
  \centering
  {\includegraphics[width=0.45\textwidth]{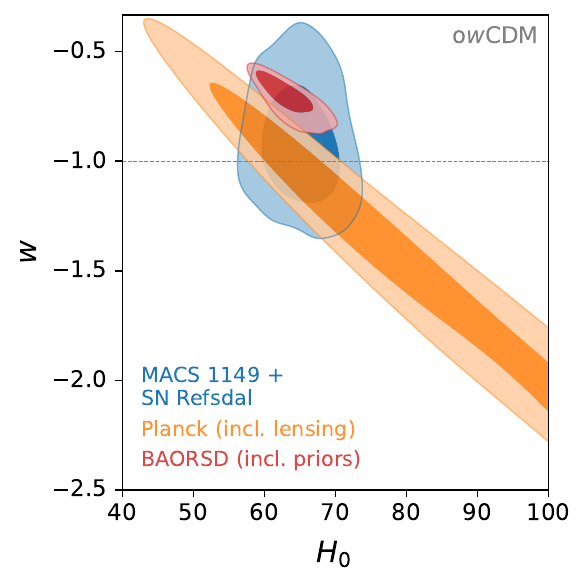}}
      \caption{Contour levels, at the 68.3\% (darker) and 95.4\% (lighter) CL, of $w$ and $H_0$ in an open $w$CDM (o$w$CDM) model for different cosmological probes: time-delay cosmography in MACS 1149 (r model; blue), Planck (orange), and BAORSD (red). The horizontal dashed line corresponds to models with $w=-1$, i.e. with a cosmological constant. BAORSD includes Gaussian priors on the values of $\omega_{\rm b}$ and $n_{\rm s}$. Details on these datasets are given in Sect. \ref{res}.}
         \label{fig:plot_2d_H0_w_owcdm}
\end{figure}

In Figs. \ref{fig:plot_1d_all_models}, \ref{fig:plot_2d_Om_H0_lcdm}, \ref{fig:plot_2d_H0_w_wcdm}, and \ref{fig:plot_2d_H0_w_owcdm}, we compare the measurements of the cosmological parameters derived from time-delay cosmography in MACS 1149 (r model) and from other probes. As parameter sampler, we employ \cosmomc\ \citep{Lewis:2002ah, Lewis:2013hha} coupled with \camb\ Boltzmann code \citep{Lewis:1999bs, Howlett:2012mh}. In what follows, we briefly outline the datasets used and the main assumptions behind each of them. In detail:
\begin{itemize}

\item \planck\ 2018 cosmic microwave background (CMB) temperature and polarization measurements, described in \cite{Aghanim:2019ame}. For this dataset, we explore the canonical parameter space, sampling both cosmological and nuisance parameters. Following \cite{Aghanim:2018eyx}, we do not vary directly the value of the Hubble parameter, but vary that of $\thetaMC$ instead (an approximation to the angular size of the sound horizon at the last scattering) and then obtain the value of $H_0$ as a derived parameter. We refer to this dataset as Planck. In Figure \ref{fig:plot_2d_H0_w_owcdm}, we also add the \Planck\ 2018 lensing potential power spectrum likelihood \citep{Aghanim:2018oex}. We label this dataset as Planck (incl. lensing). 

\item Distance ladder measurements from the SH0ES program (\citealt{rie21}) in a f$\Lambda$CDM model. We label this dataset as SH0ES.

\item Baryon Acoustic Oscillations (BAO) and Redshift-Space Distortions (RSD) measurements 
from the completed SDSS-IV eBOSS survey, as described in Table 3 of \cite{eBOSS:2020yzd}. As suggested by the authors, for this dataset we also include a BBN-inspired prior on $\omega_{\rm b}$ [i.e., $\mathcal{N}(0.0222,0.0005)$] and a prior of $n_{\rm s}$ [i.e., $\mathcal{N}(0.96,0.02)$], and we explore the same parameter space considered in the CMB analysis\footnote{The complete list of measurements, likelihood code, and settings used in \cite{eBOSS:2020yzd} are available at \url{https://github.com/evamariam/CosmoMC_SDSS2020} and at \url{https://www.sdss4.org/science/final-bao-and-rsd-measurements/}.}. We refer to this combination as BAORSD.


\item Type Ia Supernovae (SNe Ia) included in the ``Pantheon Sample'' \citep{Scolnic:2017caz}. We label this dataset as Pantheon18. Here we only explore the cosmological parameters $H_0$, $\Omega_{\rm m }$, $\Omega_\Lambda$ or $\Omega_{\rm DE}$, and $w$.

\item A collection of six gravitationally lensed quasars with measured time delays provided by the H0LiCOW collaboration \citep{won19}. In this case, we show posteriors obtained by analysing the public chains\footnote{H0LiCOW likelihood code and chains are available here: \url{https://github.com/shsuyu/H0LiCOW-public} \citep{martin_millon_2020_3633035}. These results are based on lensing distances measured using physically-motivated lens mass models by \citet{suyu+2010, suyu+2014, wong+2017, birrer+2019, chen+2019, jee+2019, rusu+2020}. }. While analyses of new lenses and data are underway with more flexible mass models in some cases \citep[e.g.,][]{shajib+2020, shajib+2022, shajib+2023, millon+2020, birrer+2020}, we use 
the publicly-available H0LiCOW chains for the f$\Lambda$CDM, o$\Lambda$CDM and f$w$CDM models, which are computed with a narrower prior on $\Omega_{\rm m}$, i.e. $[0.05,0.5]$. We refer to this dataset as H0LiCOW. 

\end{itemize}

In Fig. \ref{fig:plot_1d_all_models}, we compare the marginalised probability distribution functions of $H_{0}$, $\Omega_{\rm m}$, $\Omega_{\Lambda}$ or $\Omega_{\rm DE}$, and $w$ obtained from analysing the data of the various cosmological probes mentioned above and assuming different underlying cosmological models. Time-delay cosmography in MACS~1149 shows remarkably stable results across the models, even when relaxing the assumptions on the geometry of the Universe (as already highlighted in Fig.~\ref{fig:tri_all_models}). Despite being more sensitive within the minimal flat $\Lambda$CDM model, the other probes show probability distribution functions that vary more significantly, both in terms of centre and dispersion, in more general cosmological models. This suggests that they cannot break as efficiently as time-delay cosmography in MACS~1149 the degeneracies among the different parameters in the extended cosmological models. We notice that no other probe, without any additional priors, can measure the values of all key cosmological parameters, namely $H_{0}$, $\Omega_{\rm m}$, $\Omega_{\Lambda}$ or $\Omega_{\rm DE}$, and $w$, at the level of consistency found for time-delay cosmography in MACS~1149 in all four cosmological models considered here. In particular, the maximum differences of the parameter values obtained in the four considered cosmological models, in units of the 1$\sigma$ error of an open $w$CDM cosmological model, are approximately 0.4, 0.8, and 0.4 for $H_{0}$, $\Omega_{\Lambda}$ or $\Omega_{\rm DE}$, and $w$, respectively. As visible and already noted above, the values of $\Omega_{\rm m}$ are consistent, but their differences are statistically more significant.

In a flat $\Lambda$CDM model, the constraining power of time-delay cosmography in MACS~1149 is surpassed by both BAO (combined with redshift-space distortions) and Planck measurements (see Fig.~\ref{fig:plot_2d_Om_H0_lcdm}). In particular, the temperature and polarization data of the CMB are very sensitive to the quantity $\Omega_{\rm m}h^3$, leading to very precise estimates of both the values of $\Omega_{\rm m}$ and $H_0$. However, those constraints are strongly model-dependent \citep{Aghanim:2018eyx}. Note that BAORSD dataset is supplemented by a BBN-inspired prior that allows one to highly increase the sensitivity to $H_0$ (see \citealt{eBOSS:2020yzd} for more details).
In general, time-delay cosmography in MACS~1149 cannot measure very precisely the values of the matter and dark-energy density parameters, in particular that of $\Omega_{\rm m}$, as it is clearly visible in Fig.~\ref{fig:plot_1d_all_models}. However, relaxing the assumption of a cosmological constant and exploring the equation-of-state parameter of dark energy, $w$, Planck data lose constraining power, while time-delay cosmography in MACS~1149 is still capable of providing robust results. The same is true for most of the other cosmological probes, like H0LiCOW, for which the degeneracy between the values of $H_0$ and $w$ cannot be broken without the inclusion of additional data \citep[e.g.,][]{taubenberger+2019}. For flat $w$CDM cosmological model, the most recent results of the different probes on the plane $H_0$-$w$ are shown in Fig.~\ref{fig:plot_2d_H0_w_wcdm}. If we further relax the model, considering an open $w$CDM model, time-delay cosmography in MACS~1149 is even more competitive. As illustrated in Fig.~\ref{fig:plot_2d_H0_w_owcdm}, this is the only method that, without the addition of any external dataset or prior, is currently able to provide results that are consistent with those obtained in the other cosmological models and showing on the $H_0$-$w$ plane closed contours at the 95.4\% CL.

Interestingly, the constraints on the cosmological parameter planes from time-delay cluster lensing are oriented in such a way to be complementary to those coming from other observational techniques. Although each method probes very different physics, the results are consistent and the overall agreement on very small regions of the planes supports the validity of the concordance $w$CDM model.

We mention that the recent work by \citet{kel23b} reported a value of $H_0$
only in a f$\Lambda$CDM model in which the values of all the other cosmological parameters are fixed to $\Omega_{\rm m}=0.3$, $\Omega_{\Lambda} = 0.7$, and $w=-1$. The authors have weighted the different results, among which the non-revised ones originally provided by us, of the groups who participated in the initial analysis on the reappearance of SN Refsdal \citep{tre16}. The value of $H_0$ published by \citet{kel23b} ($64.8^{+4.4}_{-4.3}$~km~s$^{-1}$~Mpc$^{-1}$) is consistent with the blind one presented three years earlier in Table~1 of G20 (assuming a similar time-delay value of ${\rm \Delta t_{\rm SX:S1} = (375\pm10)}$ days) and with that obtained in this study (see Table~\ref{tab1}).

By considering the spectroscopically confirmed and multiply-imaged galaxies within the fields of view of five massive lens galaxy clusters and the expected number of galaxy clusters in the Vera C. Rubin Observatory Legacy Survey of Space and Time (LSST) \citep{lss09, oguri+2009}, \citet{pet20} predicts a lower limit of more than 10 SNe strongly lensed by massive galaxy clusters only. After SN Refsdal, three additional multiply-imaged and spatially-resolved SNe (named “Requiem”, “H0pe”, and “Encore”) were identified in \emph{Hubble Space Telescope} \citep{rod21} and \emph{James Webb Space Telescope} \citep{fry23} data targeting galaxy cluster fields. Cosmographic analyses through strong lensing, similar to those presented here, are currently underway in those fields. Moreover, thanks to the Sloan Digital Sky Survey observations, three \citep{ina03,ina06,dah13} out of less than a hundred multiply-imaged QSOs\footnote{http://www-utap.phys.s.u-tokyo.ac.jp/\~{}sdss/sqls/index.html} were detected behind lens galaxy groups/clusters (see e.g., \citealt{ace22b, ace22a} for some modelling results). LSST is expected to discover around 2600 strongly lensed QSOs \citep{lss09}. Thus, by simply rescaling the SDSS statistics, several tens of multiply-imaged QSOs found by LSST will be lensed by galaxy groups/clusters. Furthermore, the ongoing Euclid mission is expected to significantly increase the number of QSOs strongly lensed by galaxy clusters, providing high angular resolution imaging. These forecasts suggest that the cosmological study performed in this work will be the first among many future ones and that time-delay cluster lensing will rapidly grow in the next decade (see also \citealt{ace23,ber24}).

\section{Conclusions}
\label{sec:conclusions}

We have measured the values and the uncertainties of the Hubble constant, matter and dark-energy density parameters, and dark-energy equation-of-state parameter by means of a strong lensing analysis of the galaxy cluster MACS J1149.5+2223. We have exploited high-quality photometric and spectroscopic data for a large number of multiple images at different redshifts and the new measured time delays of the multiple images of SN Refsdal. Compared to other standard cosmological probes, we have shown that this method can provide very stable results, depending only very mildly on the underlying cosmological model and on the lens modelling details. We have found only small differences in the probability distribution functions of the key cosmological parameters by extending the cosmological model from a flat $\Lambda$CDM to an open $w$CDM model. We have also demonstrated that the so-called ``mass-sheet'' and ``mass-slope'' degeneracies are substantially mitigated by the presence of many multiple images, observed over relatively large ranges of projected distances from the cluster centre and redshifts. In particular, $H_0$ constraints only vary within 1.4~km~s$^{-1}$~Mpc$^{-1}$ for all the lensing models we have considered. Without using any priors from previous cosmological experiments, we have obtained that, in an open $w$CDM model, this novel technique offers results that are superior to those of the most recent CMB and BAO data, producing combined statistical and systematic relative errors (at the 68.3\% CL) of approximately 6\% and 21\% for $H_0$ and $w$, respectively. We expect that time-delay cluster lensing will become a new accurate and precise cosmological probe, thanks also to the many cluster strong lenses that will be discovered in forthcoming deep and wide surveys.

\begin{acknowledgements}
C.G. and P.R. acknowledge support through grants MIUR2017 WSCC32 and MIUR2020 SKSTHZ.
L.P.~acknowledges the financial support from the INFN InDark project and from the COSMOS network (www.cosmosnet.it) through the ASI (Italian Space Agency) Grants 2016-24-H.0 and 2016-24-H.1-2018.
S.H.S.~thanks the Max Planck Society for support through the Max Planck Research Group and Fellowship, and the European Research Council (ERC)
for support under the European Union’s Horizon 2020 research and innovation programme (LENSNOVA: grant agreement No 771776).
This work is based in large part on data collected in service mode at ESO VLT, through the Director's Discretionary Time Programme 294.A-5032.
The CLASH Multi-Cycle Treasury Program is based on observations made with the NASA/ESA {\it Hubble Space Telescope}. The Space Telescope Science Institute is operated by the Association of Universities for Research in Astronomy, Inc., under NASA contract NAS 5-26555. ACS was developed under NASA Contract NAS 5-32864. 

We acknowledge the use of \texttt{numpy} \citep{Harris:2020xlr}, \texttt{matplotlib} \citep{Hunter:2007ouj} and \texttt{getdist} \citep{Lewis:2019xzd} software packages, and the use of computing facilities at CINECA.

\end{acknowledgements}

%
%

\bibliographystyle{aa} 
\bibliography{refs} 

\begin{appendix} 

\end{appendix}

\end{document}